\documentclass[aps,prd,eqsecnum,preprint,tightenlines,nofootinbib]{revtex4-2}
\usepackage{graphicx,latexsym}

\def\bea{\begin{eqnarray}}
\def\eea{\end{eqnarray}}
\def\be{\begin{equation}}
\def\ee{\end{equation}}

\newcommand{\prtl}[2]{\frac{\partial #1}{\partial #2}}

\begin{document}

\title{Schr\"odinger-Newton solitons with axial symmetry
}
\author{A. Flores}
\author{C. Stegner}
\author{S.S.~Chabysheva}
\affiliation{Department of Physics, University of Idaho, Moscow, Idaho 83844 USA}
\author{J.R.~Hiller}
\affiliation{Department of Physics, University of Idaho, Moscow, Idaho 83844 USA}
\affiliation{Department of Physics and Astronomy,
University of Minnesota-Duluth,
Duluth, Minnesota 55812 USA}

\date{\today}

\begin{abstract}

We solve the Schr\"odinger-Newton problem of Newtonian
gravity coupled to a nonrelativistic scalar particle
for solutions with axial symmetry.  The gravitational
potential is driven by a mass density assumed to be 
proportional to the probability density of the scalar.
Unlike related calculations for condensates of
ultralight dark matter or boson stars, no assumption 
of spherical symmetry is made for the 
effective gravitational potential.  Instead, the potential has only axial symmetry,
consistent with the axial symmetry of the particle's probability
density for eigenstates of $L_z$.  With total angular
momentum no longer a good quantum number, there are
in general contributions from a range of partial waves.
This permits us to study the partial-wave content of self-consistent
solutions of the Schr\"odinger-Newton system.

\end{abstract}

\maketitle

\section{Introduction}  \label{sec:Introduction}

We consider the states of a quantum particle bound in a gravitational
well of its own making.  The mass of the particle times its
probability distribution provides the source for the Poisson
equation that determines the Newtonian gravitational potential.
The Schr\"odinger equation is then solved for this potential, to
find the particle's wave function.  The self-consistent solution
for a chosen energy level in the Schr\"odinger spectrum is then a
soliton, a solitary wave.  Previously, we~\cite{Taylor} and
others~\cite{Ruffini,Moroz,Bernstein,Tod,HarrisonMorozTod,Zagorac,AlvarezRios}
considered solitons with spherical symmetry that are
eigenstates of $L^2$ and $L_z$; here we extend this to axial 
symmetry and eigenstates of $L_z$ alone, with a more general
treatment of this case than previously 
considered~\cite{Schupp,Silveira,Yoshida,Schunck,Harrison,%
Guzman1,Alcubierre,Guzman2,Jaramillo,Dmitriev,Nambo}.

There are many applications for such solitons in astrophysics
and cosmology.  Ultralight dark matter~\cite{Schive,Guth,Chavanis,Zagorac,AlvarezRios}
may form a Bose-Einstein condensate for which a Hartree-type
analysis yields a single-particle wave function bound in an
effective gravitational potential generated by all the 
other dark-matter particles.  For nonrelativistic systems,
this is mathematically equivalent to the Schr\"odinger-Newton
system of equations as if each particle was bound by its
own gravitational potential.  The only quantitative
difference is that in the case of the condensate,
the effective potential is multiplied by the square of the number
of particles.

The same mathematics arises in the case of
boson stars~\cite{Kaup,Ruffini,Ferrell,Schupp,Silveira,Yoshida,Schunck,Guzman1,Guzman2,Dmitriev}.  
Again, individual particles experience an effective potential derived
from the mass distributions of all the others.  This
is extended to $l$-boson stars~\cite{Alcubierre,Jaramillo,Nambo}, where states with nonzero
angular momentum $l$ are filled in such a way that each
angular momentum projection is equally likely, which
guarantees a spherically symmetric source for the
gravitational field.  The gravitational potential
is then spherically symmetric.\footnote{In \protect\cite{Guzman2}, 
spherical symmetry of the effective potential is implicit.}
Axially symmetric solutions of the quantum eigenvalue 
problem are then obtained for this spherically symmetric 
potential.

On a more fundamental level, the Schr\"odinger-Newton problem also arose 
in the context of wave function collapse~\cite{Diosi,Penrose,Bahrami}.  If the state
of a system involves a superposition of two (or more) probability
amplitudes with spatial separation, the gravitational interaction
between the associated mass distributions should result in decay
of the superposition over a time scale inversely related to the
gravitational self-energy~\cite{DiMauro}.  Such a superposition
might decay into a Schr\"odinger-Newton soliton, and various
calculations of these solitons have been 
done~\cite{Moroz,Bernstein,Tod,HarrisonMorozTod,Harrison},
primarily with an assumption of spherical symmetry.

The nonrelativistic problem is also obtained beginning with a semiclassical
treatment of gravity~\cite{Bahrami}, where the gravitational field
is the solution to the equations of general relativity but the 
source is a stress-energy tensor computed from quantum mechanical
amplitudes.  For a scalar particle one can solve for the 
amplitude from the Klein-Gordon equation in curved spacetime~\cite{Lehn}, to
obtain a self-consistent Einstein-Klein-Gordon soliton~\cite{Taylor}.
The nonrelativistic reduction of semiclassical gravity yields again
the Schr\"odinger-Newton problem~\cite{Giulini,Brizuela}.

In this paper we consider the Schr\"odinger-Newton problem
in the more general situation of an axially symmetric gravitational
potential. The wave functions are no longer necessarily eigenfunctions
of $L^2$ but only of $L_z$.  However, their simple $e^{im\phi}$ dependence on the
azimuthal angle results in a probability density
that is independent of the azimuthal angle.\footnote{Harrison~\protect\cite{Harrison}
assumed wave functions independent of the azimuthal angle, but
this is unnecessarily restrictive.}  The source for
the gravitational potential is then axially 
symmetric, making the symmetry of the solutions
self-consistent.  Spherically symmetric solutions of this system are,
of course, special cases, which we do reproduce.

With the azimuthal dependence trivial, the problem
becomes two dimensional.  We approach it in two ways.
The first is as a set of two-dimensional partial
differential equations, the Schr\"odinger equation
and the Poisson equation.\footnote{Schupp and van der Bij~\protect\cite{Schupp}
approach the problem in this way; we agree with their only
nonspherical result.}  The second uses partial-wave 
expansions of the wave function and potential.
For the former approach, we find serious
problems with convergence of the iteration for 
self-consistency except for the ground state, 
unless the initial guess is very close to the final
result.  Convergence in the case of the partial-wave
expansions is quite rapid, requiring on the order of
twenty iterations to achieve accuracy consistent 
with the underlying numerical methods for the
differential equations.  Results from this method
are confirmed by using the first method starting
from interpolations of the solution from the second method, which
is then usually sufficient to obtain convergence.

Our approach differs from earlier work with 
partial waves~\cite{Silveira}.  There the sum
was restricted to a single partial wave that was
assumed to be dominant.  We instead sum over all
partial waves that make any significant contribution
and tabulate the range of contributions.
In doing so, we found a case where two partial
waves were equally important.

The poor convergence of the first method appears to
be due to the fact that a self-consistent solution
can exist anywhere in space; there is no pre-determined 
location for the ``center'' of the soliton.  
Within the restriction to cylindrical coordinates,
and with the assumption of cylindrical symmetry with
respect to these coordinates, the soliton is still
able to appear anywhere along the symmetry axis.
In any excited state there are secondary peaks in the
wave function which can confuse the numerical 
algorithm into seeing them as a central source,
and the iteration seeks to find a solution at
each new location along the axis of symmetry.
The partial-wave expansion defeats this because
any secondary peaks in radial wave functions 
correspond to spherical shells rather than new
centers.

A mathematical description of the Schr\"odinger-Newton problem is
given in Sec.~\ref{sec:SNproblem}, where we also
describe our methods of solution.  Results
are presented in Sec.~\ref{sec:results} and summarized
in Sec.~\ref{sec:summary}.

\section{Schr\"odinger-Newton problem}  \label{sec:SNproblem}

Given the Hamiltonian $H=-\frac{1}{2m}\nabla^2+V$, with $V=m\Phi$
the potential energy in a gravitational potential $\Phi$, the coupled system is
\be
H\psi=E\psi,\;\;
\nabla^2\Phi=4\pi Gm|\psi|^2.
\ee
Here $m|\psi|^2$ is taken to be the mass density.
We assume axial symmetry so that $\Phi$ is independent
of the azimuthal angle $\phi$ and $H$ commutes with $L_z$.
The wave function is then an eigenstate of $L_z$ and 
takes the form $\frac{1}{\sqrt{2\pi}}R_{m_l}(\rho,z)e^{im_l\phi}$,
with $z$ the symmetry axis and $\rho$ the radial distance from it.
The mass density is just $\frac{m}{2\pi}|R_{m_l}(\rho,z)|^2$ and
also independent of azimuthal angle.

We also assume that the gravitational potential $\Phi(\rho,z)$ is even
with respect to reflection in $z$.  The Hamiltonian is then 
invariant in $z$ parity, and the eigenfunctions $R_{m_l}$
can be chosen even or odd.  This makes the source
of the gravitational potential even and self-consistent with the chosen
symmetry of the potential.

Solutions with spherical symmetry are recovered as a subset of
those solutions with $m_l=0$.  We use this as a partial check
on the calculations.  A second check is to compare results from solving the system in
two ways, one using cylindrical coordinates $\rho$ and $z$, as already introduced,
and the other using spherical coordinates $r$ and $\theta$, with
$z=r\cos\theta$ and $\rho=r\sin\theta$ as usual.

\subsection{Cylindrical coordinates}  \label{sec:cylindrical}

In cylindrical coordinates, the Schr\"odinger equation reduces to
\be
-\frac{1}{2m}\left[\frac{1}{\rho}\prtl{}{\rho}\left(\rho\prtl{R_{m_l}}{\rho}\right)
     -\frac{m_l^2}{\rho^2}R_{m_l}+\prtl{^2R_{m_l}}{z^2}\right]+V_{m_l}(\rho,z)R_{m_l}=E_{m_l}R_{m_l},
\ee
and, when written in terms of the potential energy $V=m\Phi$,
the Poisson equation for Newtonian gravity becomes
\be
\frac{1}{\rho}\prtl{}{\rho}\left(\rho\prtl{V_{m_l}}{\rho}\right)+\prtl{^2V_{m_l}}{z^2}
   =2 G m^2|R_{m_l}|^2.
\ee
The wave function is normalized as
\be
1=\int \rho d\rho dz d\phi|\psi|^2=\int\rho d\rho dz |R_{m_l}|^2,
\ee
so that the total mass is $m$ with a density of $m|\psi|^2$.

To have a dimensionless representation, as discussed elsewhere~\cite{Lehn,Taylor},
we rescale lengths by the gravitational Bohr radius $a=1/Gm^3$,
energies by $G^2m^5$, with $\epsilon_{m_l}\equiv E_{m_l}/G^2m^5$,
and the wave function by $a^{-3/2}=(Gm^3)^{3/2}$ to obtain
in the same notation
\be
-\frac{1}{2}\left[\frac{1}{\rho}\prtl{}{\rho}\left(\rho\prtl{R_{m_l}}{\rho}\right)
     -\frac{m_l^2}{\rho^2}R_{m_l}+\prtl{^2R_{m_l}}{z^2}\right]+V_{m_l}(\rho,z)R_{m_l}=\epsilon_{m_l}R_{m_l}
\ee
and
\be
\frac{1}{\rho}\prtl{}{\rho}\left(\rho\prtl{V_{m_l}}{\rho}\right)+\prtl{^2V_{m_l}}{z^2}=2|R_{m_l}|^2.
\ee
Given all the rescalings, the normalization expression is invariant,
\be
1=\int\rho d\rho dz |R_{m_l}|^2.
\ee

It is also convenient to introduce the reduced wave function $u_{m_l}\equiv \sqrt{\rho}R_{m_l}$
for which the 2D Schr\"odinger equation becomes
\be
-\frac{1}{2}\left[\prtl{^2u_{m_l}}{\rho^2}
     -\frac{m_l^2-1/4}{\rho^2}u_{m_l}+\prtl{^2u_{m_l}}{z^2}\right]+V_{m_l}(\rho,z)u_{m_l}=\epsilon_{m_l}u_{m_l}.
\ee
However, for $m_l=0$ the boundary condition $u_{m_l}(\rho=0,z)=0$ is difficult to maintain 
numerically; cancellations between the $1/4\rho^2$ term and finite-difference expressions
for the partial derivatives with respect to $\rho$ are imprecise near $\rho=0$.  In this case,
the original equation for $R_0$ is the better route despite the more complicated $\rho$ 
derivatives.

To solve each of the equations in the system, we apply finite difference approximations for all
derivatives on a finite grid and study the dependence on grid spacing and grid size.
The finite-difference representation of the Schr\"odinger equation is solved as a 
matrix eigenvalue problem and that of the Poisson equation is solved with the usual
successive-over-relaxation (SOR) method.
The coupled system is then solved iteratively, starting from an initial guess for
the wave function, until both $R_{m_l}$ and $V_{m_l}$ converge to consistent
solutions.  This requires a choice of energy level from the multiple solutions
of the Schr\"odinger eigenvalue problem; each generates its own unique
gravitational potential because each energy level has a different 
wave function and therefore a different probability distribution.

One aspect that is important to note is that,
although the grid is chosen large enough for the wave function to be effectively
zero at the boundary, the potential $V$ does not fall nearly fast enough to be well
approximated by zero there.  Instead, we represent $V$ by a truncated multipole
expansion computed from the mass distribution under the assumption that all the
mass is contained within the grid.  Once the moments of the mass distribution are
computed, the value of $V$ along the boundary can be estimated.
In our rescaled units, the multipole expansion is
\be
V_{m_l}(\rho,z)=-\sum_l\frac{Q_{lm_l}}{(\sqrt{\rho^2+z^2})^{l+1}}P_l(z/\sqrt{\rho^2+z^2}),
\ee
with moments
\be  \label{eq:Qlm}
Q_{lm_l}=\int (\sqrt{\rho^2+z^2})^l \rho d\rho dz  P_l(z/\sqrt{\rho^2+z^2})|R_{m_l}|^2
\ee
and $P_l$ the Legendre polynomial of order $l$. The leading term
in $V_{m_l}$ is $-1/r$, the 1 being the normalization of $R_{m_l}$.
For odd $l$, the moments are zero because $|R_{m_l}|^2$ is even in $z$.
The other moments are computed numerically with trapezoidal
approximations to the integrals.

As discussed in the Introduction, the convergence of the system iteration
is problematic.  Except for the spherically symmetric ground state, the 
iteration typically requires a very good estimate for the initial guess.
We therefore use the formulation in cylindrical coordinates as a check
on our method for spherical coordinates.

\subsection{Spherical coordinates}  \label{sec:spherical}

In terms of spherical coordinates, the natural approach is
one of partial waves.\footnote{As mentioned in the Introduction,
this approach was used in \protect\cite{Silveira}, but the sum over 
partial waves was restricted to a single partial wave, 
assumed to be dominant.}  By expanding both the wave function
and the potential in terms of spherical harmonics, one can
obtain coupled sets of ordinary differential equations for
the expansion coefficients, which are functions of the 
radial distance $r$.  To have a finite set of equations,
the expansion is truncated at a maximum value $l_{\rm max}$ for 
the angular momentum quantum number $l$;
of course, the dependence on $l_{\rm max}$ must be checked.  The
coefficient functions for different $l$ are coupled in 
general, because the potential is only axially 
symmetric, not spherically symmetric.

The partial-wave expansions are
\be   \label{eq:partialwaves}
\psi_{m_l}(r,\theta,\phi)=\sum_{l=|m_l|}^{l_{\rm max}}\frac{u_{lm_l}(r)}{r}Y_{lm_l}(\theta,\phi),\;\;
V_{m_l}(r,\theta)=\sum_{l=0}^{l_{\rm max}}\frac{v_{lm_l}(r)}{r}Y_{l0}(\theta).
\ee
For $\psi$, only $l$ values with $l-|m_{l}|$ even (odd) contribute
to states even (odd) in $z=r\cos\theta$, as determined by the associated 
Legendre functions in the $Y_{lm_l}$.  For $V$, only $Y_{l0}$ with
$l$ even contribute because $V$ is independent of $\phi$ and even in $z$.

With the definition of the overlap integrals
\be
C_{l,l'l''}^{m_l}\equiv \int d\Omega\, Y_{lm_l}^* Y_{l'0} Y_{l''m_l},
\ee
the coupled systems are
\be
-\frac12\frac{d^2 u_{lm_l}}{dr^2}+\frac{l(l+1)}{2r^2}u_{lm_l}
+\frac{1}{r}\sum_{l',l''}^{l_{\rm max}}C_{l,l'l''}^{m_l}v_{l'm_l}u_{l''m_l}
=\epsilon_{m_l}u_{lm_l}
\ee
and
\be
\frac{d^2v_{lm_l}}{dr^2}-\frac{l(l+1)}{r^2}v_{lm_l}
=\frac{4\pi}{r}\sum_{l'l''}C_{l',ll''}^{m_l*}u_{l'm_l}u_{l''m_l}.
\ee
Again a grid is introduced, though only one-dimensional in this case,
out to a finite range of $r_{\rm max}$, and the equations are
replaced by finite difference approximations.  The system for
the $u_{lm_l}$ is solved as a matrix eigenvalue problem 
for the chosen level and $z$ parity at a fixed value of $m_l$.
The system for the $v_{lm_l}$ is not coupled between different $l$
and each (inhomogeneous) finite-difference equation is solved as 
a linear system. The overlap integrals are computed from Clebsch-Gordan
coefficients with use of the Python procedure sympy.physics.quantum.cg;
the connection is 
\be
C_{l,l'l''}^{m_l}=\sqrt{\frac{(2l'+1)(2l''+1)}{4\pi(2l+1)}}
\langle l'l''00|(l'l'')l0\rangle \langle l'l''0m_l|(l'l'')lm_l\rangle.
\ee

The boundary conditions for these systems are 
\be
u_{lm_l}(0)=0,\;\;
u_{lm_l}(r_{\rm max})=0
\ee
and
\be
v_{lm_l}(0)=0,\;\;
v_{lm_l}(r_{\rm max})=-\sqrt{\frac{4\pi}{2l+1}}\frac{Q_{lm_l}}{r_{\rm max}^l},
\ee
with $Q_{lm_l}$ defined in (\ref{eq:Qlm}).
As in the case of cylindrical coordinates, the boundary condition
for $v_{lm_l}$ at $r_{\rm max}$ comes from a comparison with
the multipole expansion in the region outside the source. 
In spherical coordinates, this expansion is
\be
V_{m_l}(r,\theta)=-\sum_l\frac{Q_{lm_l}}{r^{l+1}}P_l(\cos\theta),
\ee
and the moments are written as
\be
Q_{lm_l}=\int r^{l+2} dr d(\cos\theta) d\phi P_l(\cos\theta)|\psi|^2.
\ee
Given the partial-wave expansion for $\psi$ in (\ref{eq:partialwaves}), 
the moments reduce to
\be
Q_{lm_l}=\sqrt{\frac{4\pi}{2l+1}}\sum_{l',l''=|m_l|}^{l_{\rm max}} C_{l',ll''}^{m_l}\int r^l\,dr u_{l'm_l} u_{l''m_l}.
\ee
The leading factors depending on $4\pi$ and $2l+1$ are due to
the connection $Y_{l0}=\sqrt{\frac{2l+1}{4\pi}}P_l$.

Because each partial wave is orthogonal, the normalization condition
for the radial wave functions is just a simple sum of integrals:
\be
1=\sum_{l=|m_l|}^{l_{\rm max}} \int dr |u_{lm_l}|^2
\ee
This determines an overall normalization constant.  The relative
normalization for each $l$ is determined by the solution of
the eigenvalue problem.

The combination of the partial-wave equations for the radial
wave functions and the potential energy functions is again
solved iteratively, beginning with a simple Gaussian as the
initial guess for the lowest partial wave.  Convergence to
four significant figures typically requires twenty iterations.

\section{Results}  \label{sec:results}

\begin{figure}[ht]
\vspace{0.2in}
\begin{tabular}{cc}
\includegraphics[width=8cm]{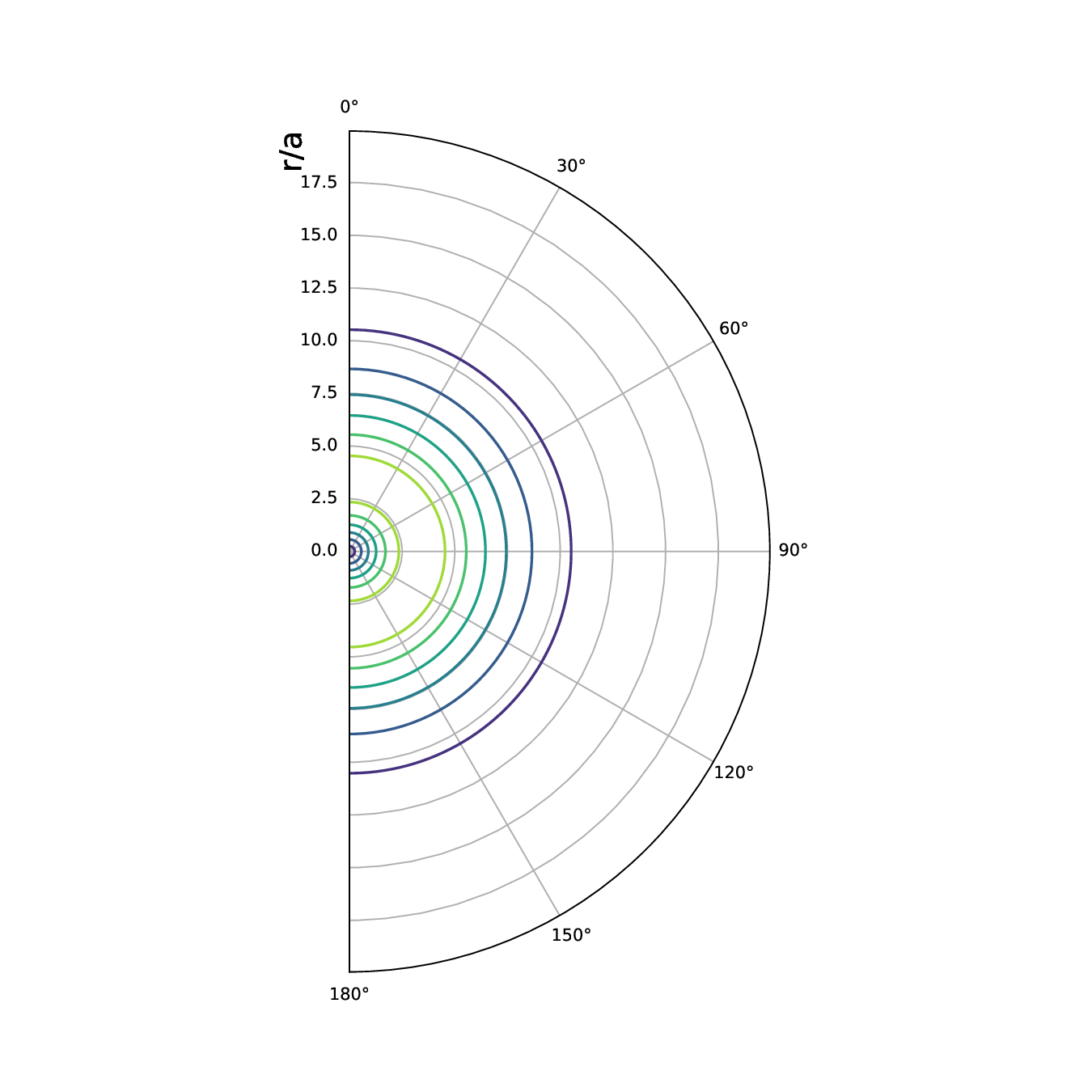} & 
\includegraphics[width=8cm]{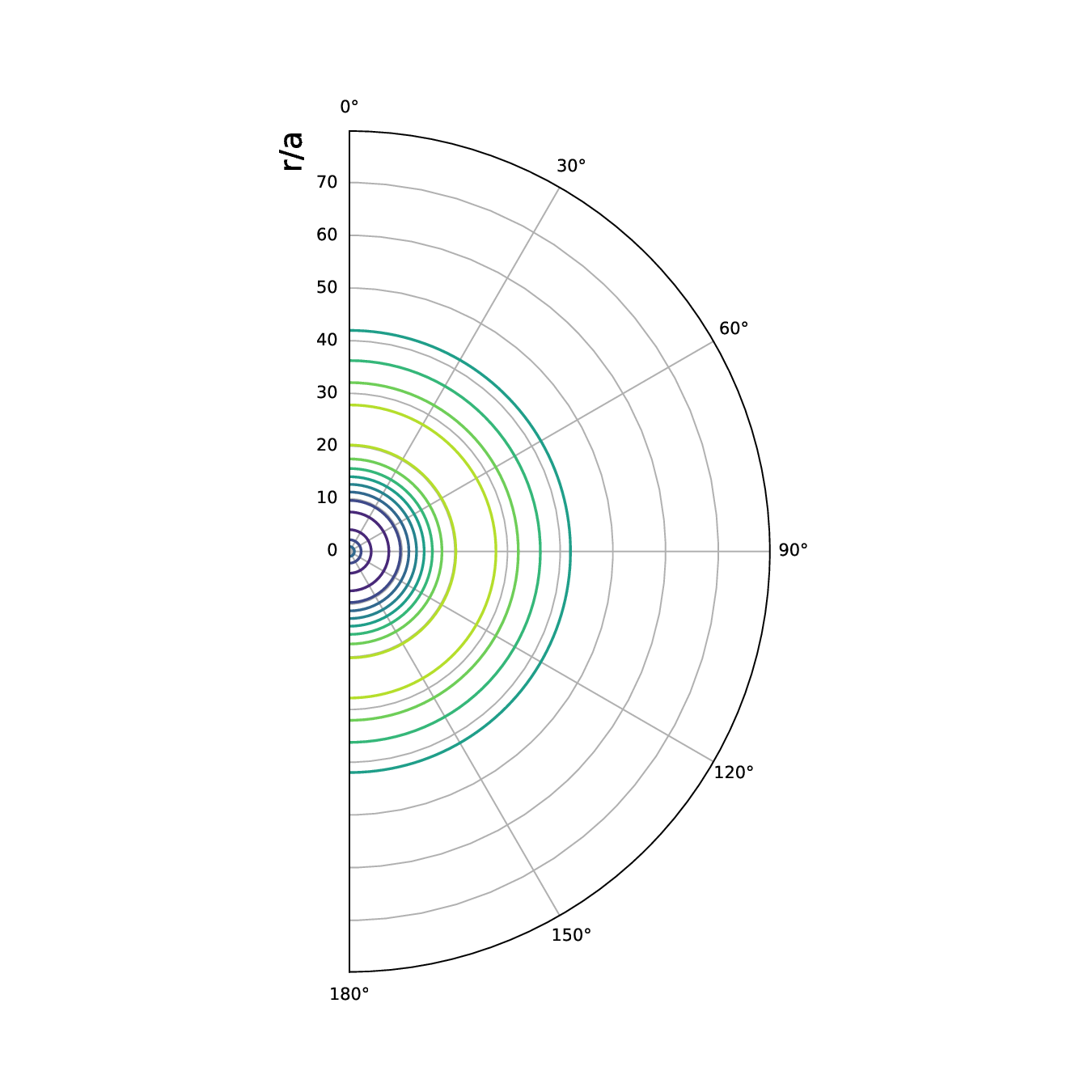} \\
(a) & (b) \\
\includegraphics[width=8cm]{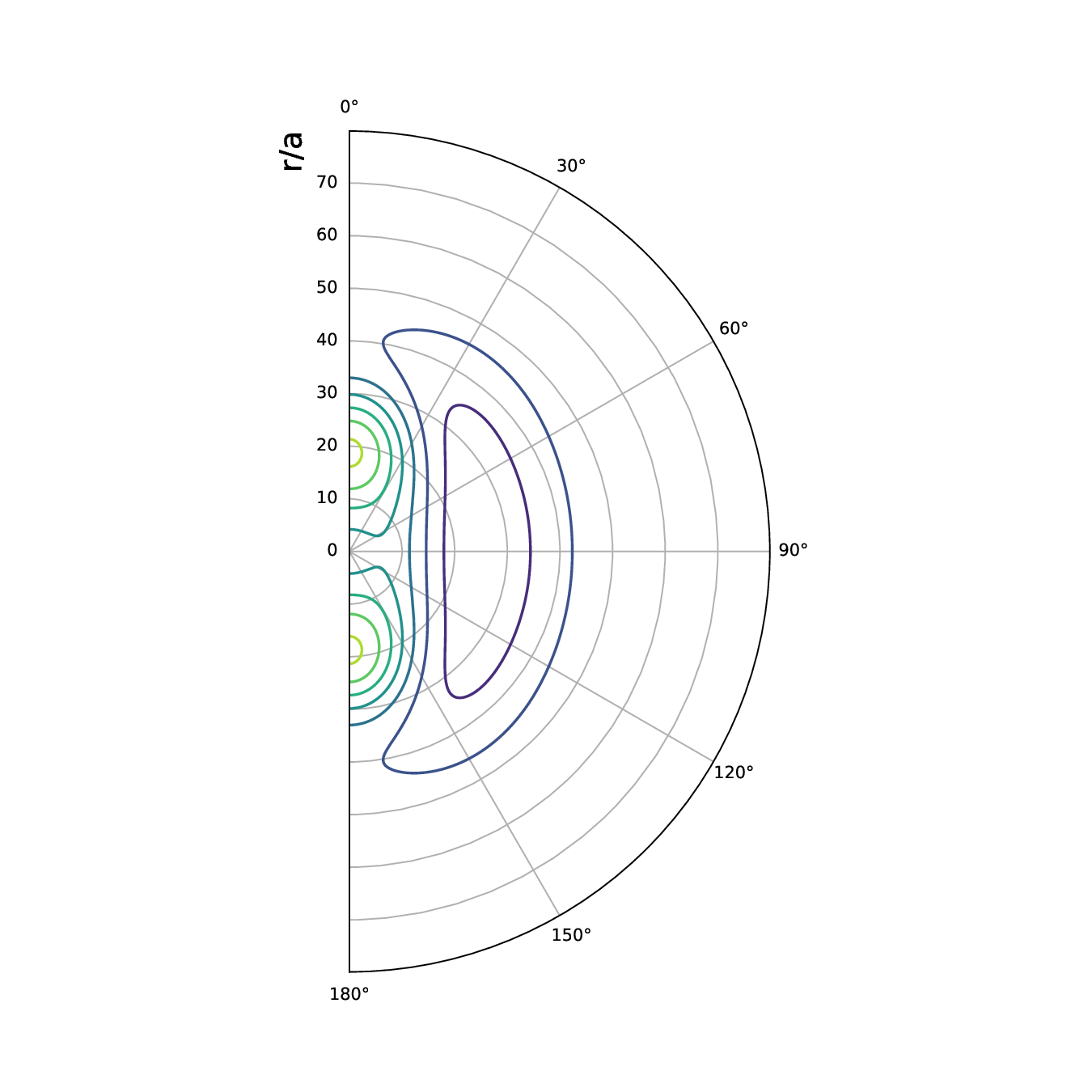} & 
\includegraphics[width=8cm]{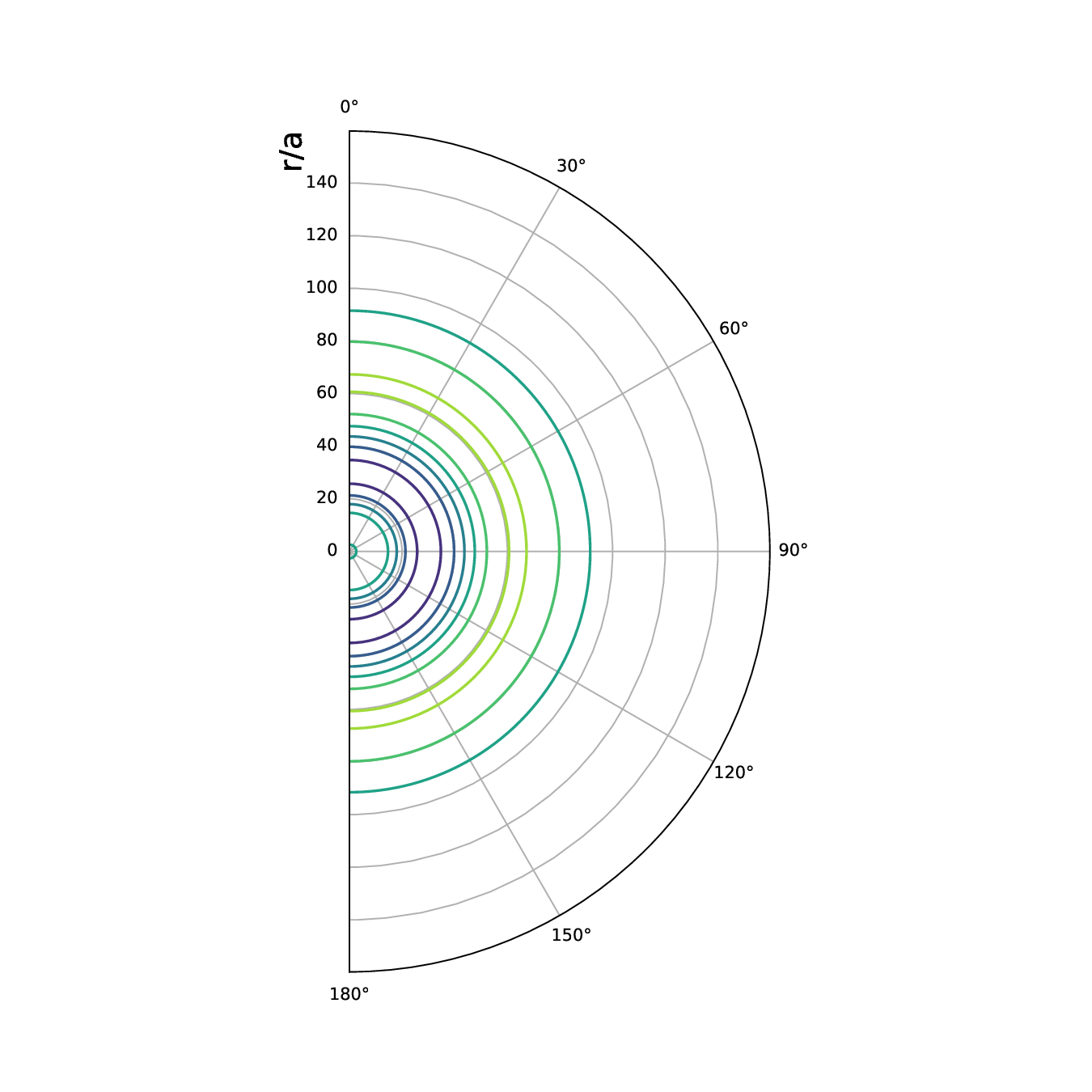} \\
(c) & (d)
\end{tabular}
\caption{\label{fig:ml0even}
Probability densities for the four lowest-lying states
with $m_l=0$ and even $z$ parity.  The associated
eigenenergies in units of $G^2m^5$ are 
(a) -0.1628,
(b) -0.03082,
(c) -0.0252, and
(d) -0.01254.  Only (c) is not spherically symmetric.
}
\end{figure}

\begin{figure}[ht]
\vspace{0.2in}
\begin{tabular}{cc}
\includegraphics[width=8cm]{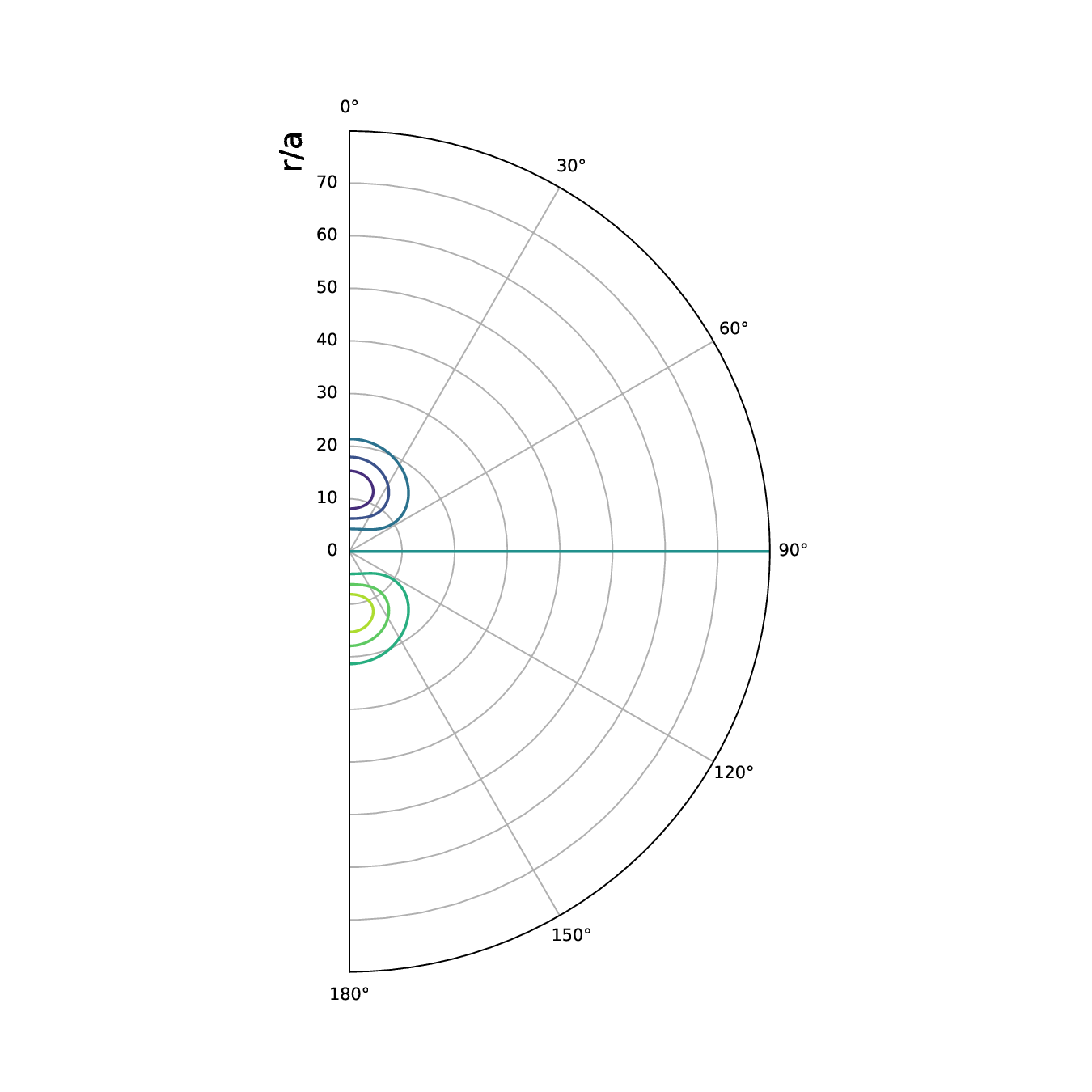} & 
\includegraphics[width=8cm]{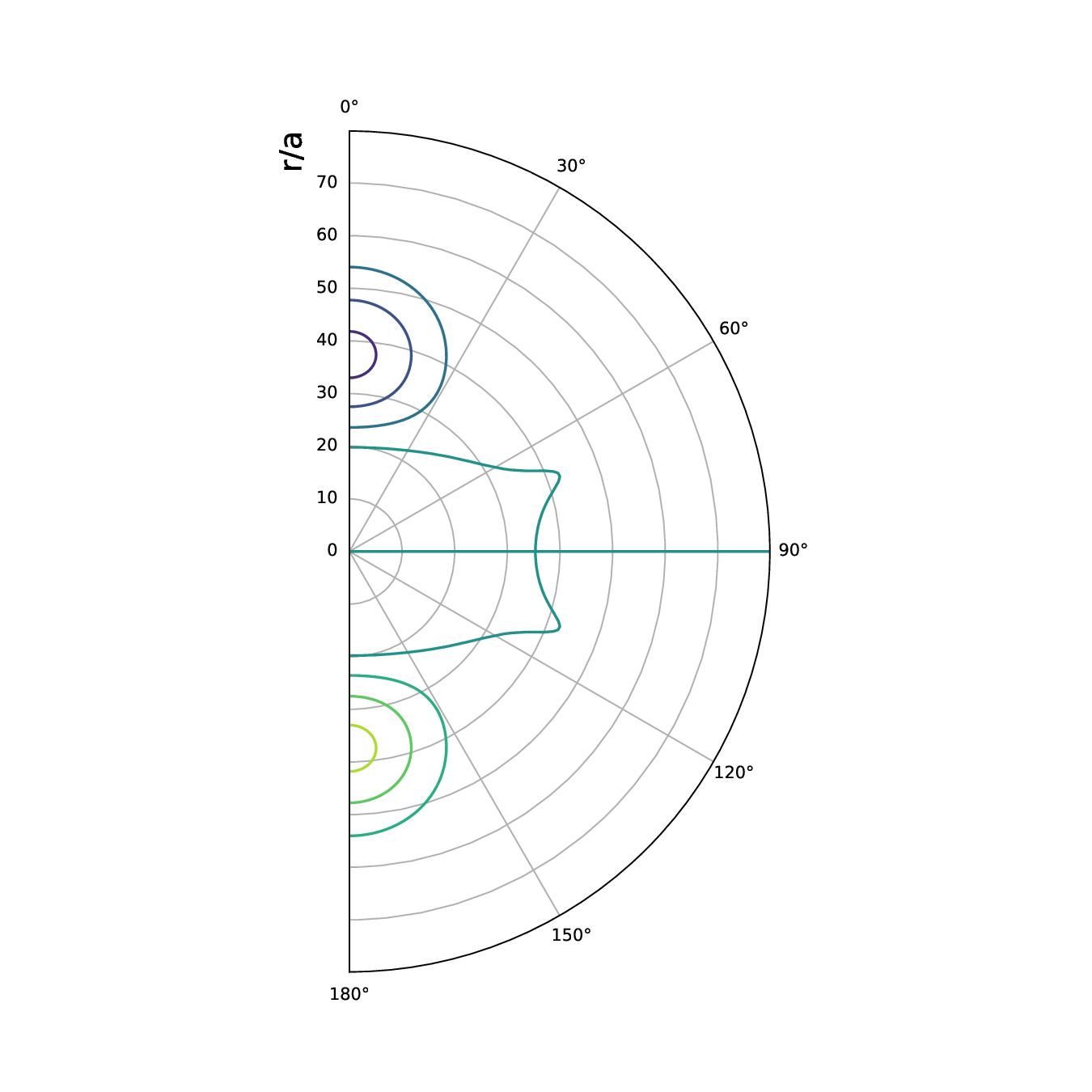} \\
(a) & (b) \\
\multicolumn{2}{c}{\includegraphics[width=8cm]{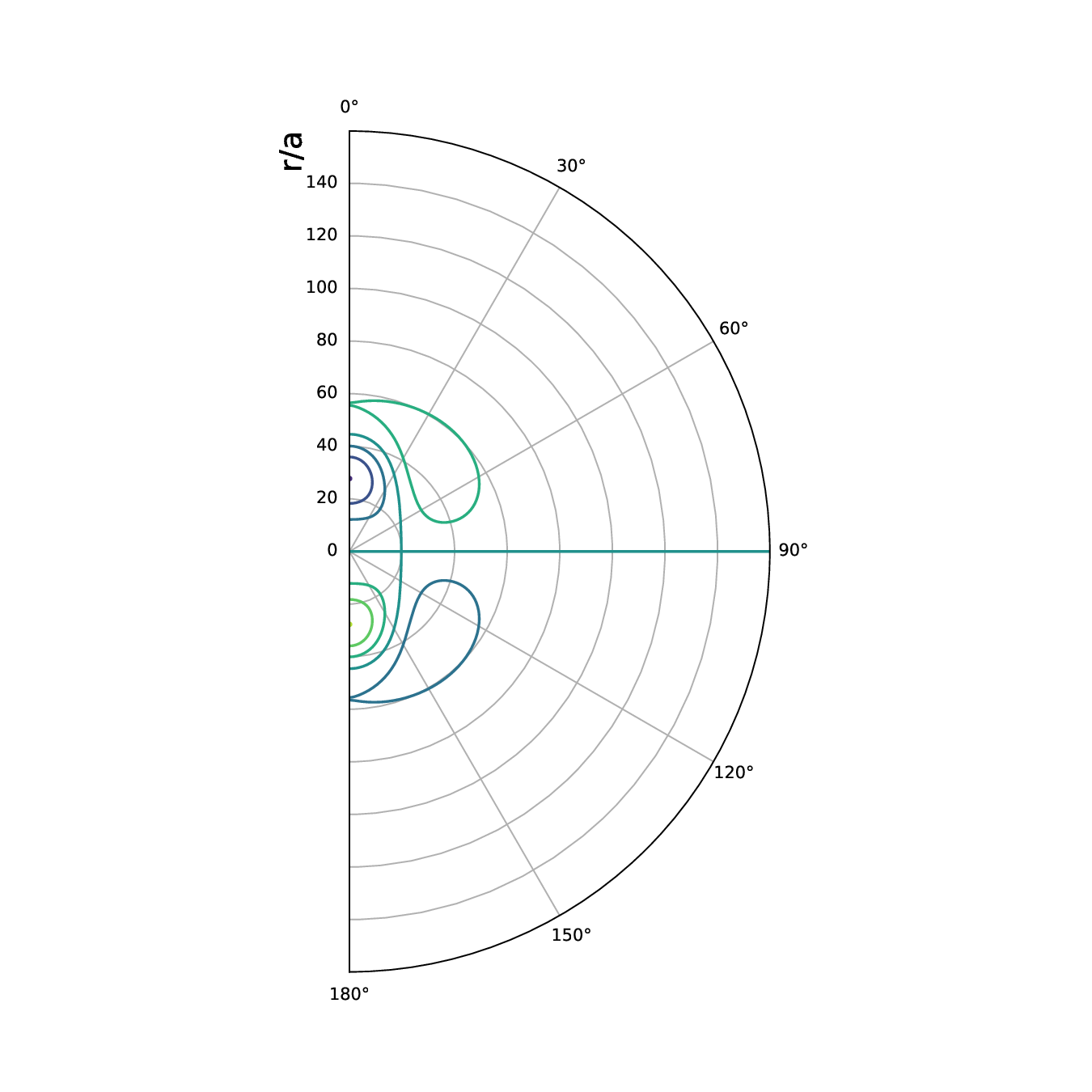}} \\
\multicolumn{2}{c}{(c)}
\end{tabular}
\caption{\label{fig:ml0odd}
Probability densities for the three lowest-lying states
with $m_l=0$ and odd $z$ parity.  The associated
eigenenergies in units of $G^2m^5$ are 
(a) -0.06894,
(b) -0.0274, and
(c) -0.0169.
}
\end{figure}

\begin{figure}[ht]
\vspace{0.2in}
\begin{tabular}{cc}
\includegraphics[width=8cm]{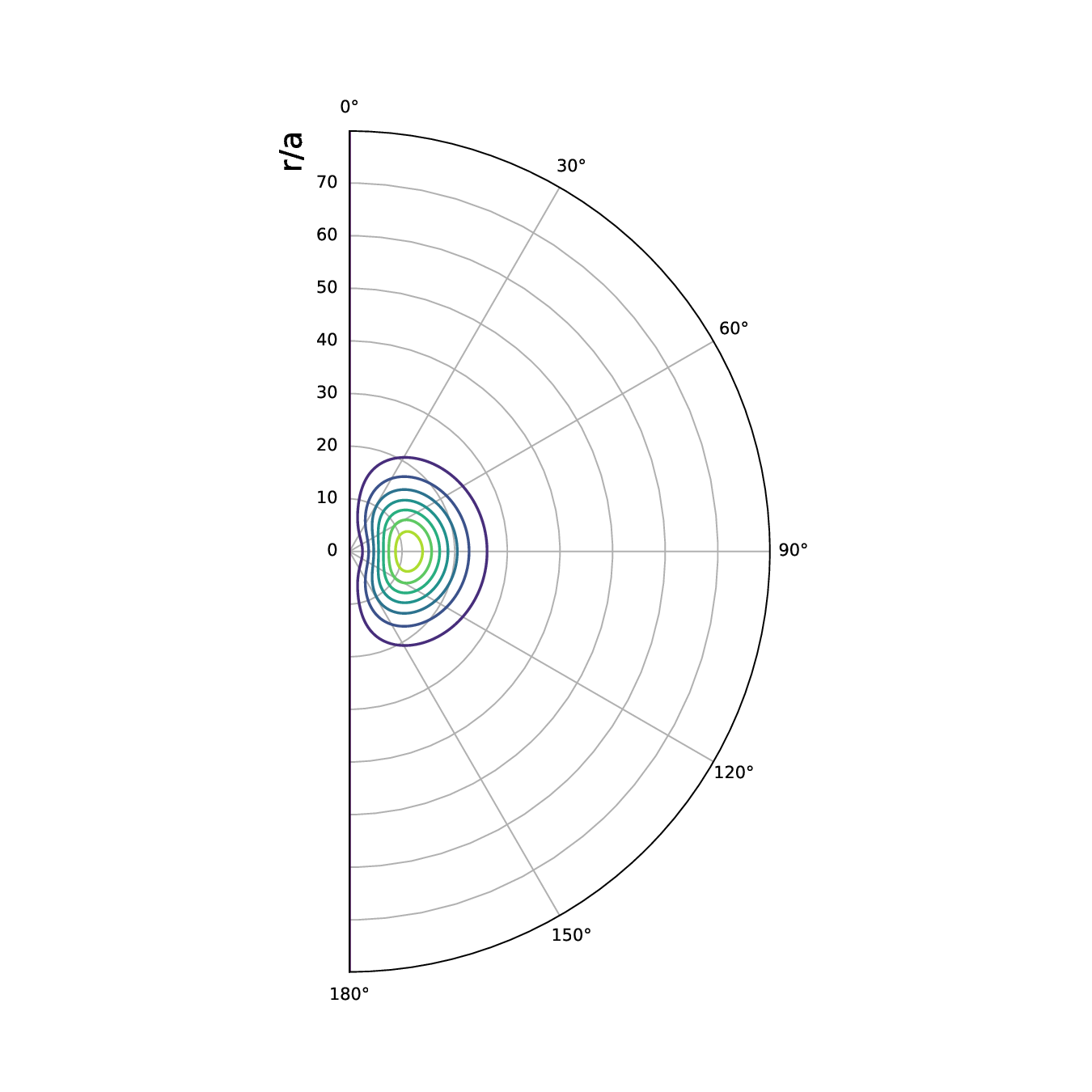} & 
\includegraphics[width=8cm]{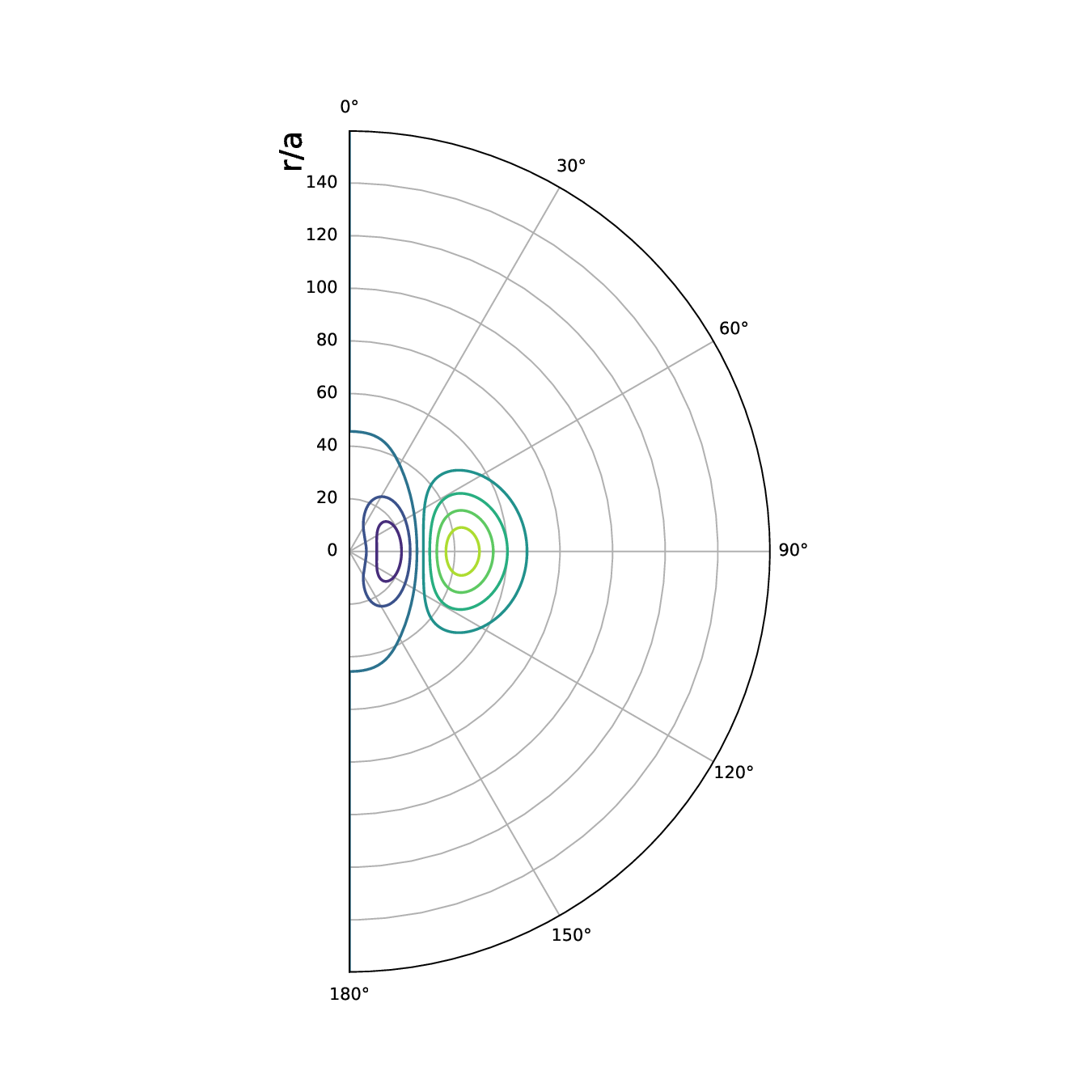} \\
(a) & (b) \\
\includegraphics[width=8cm]{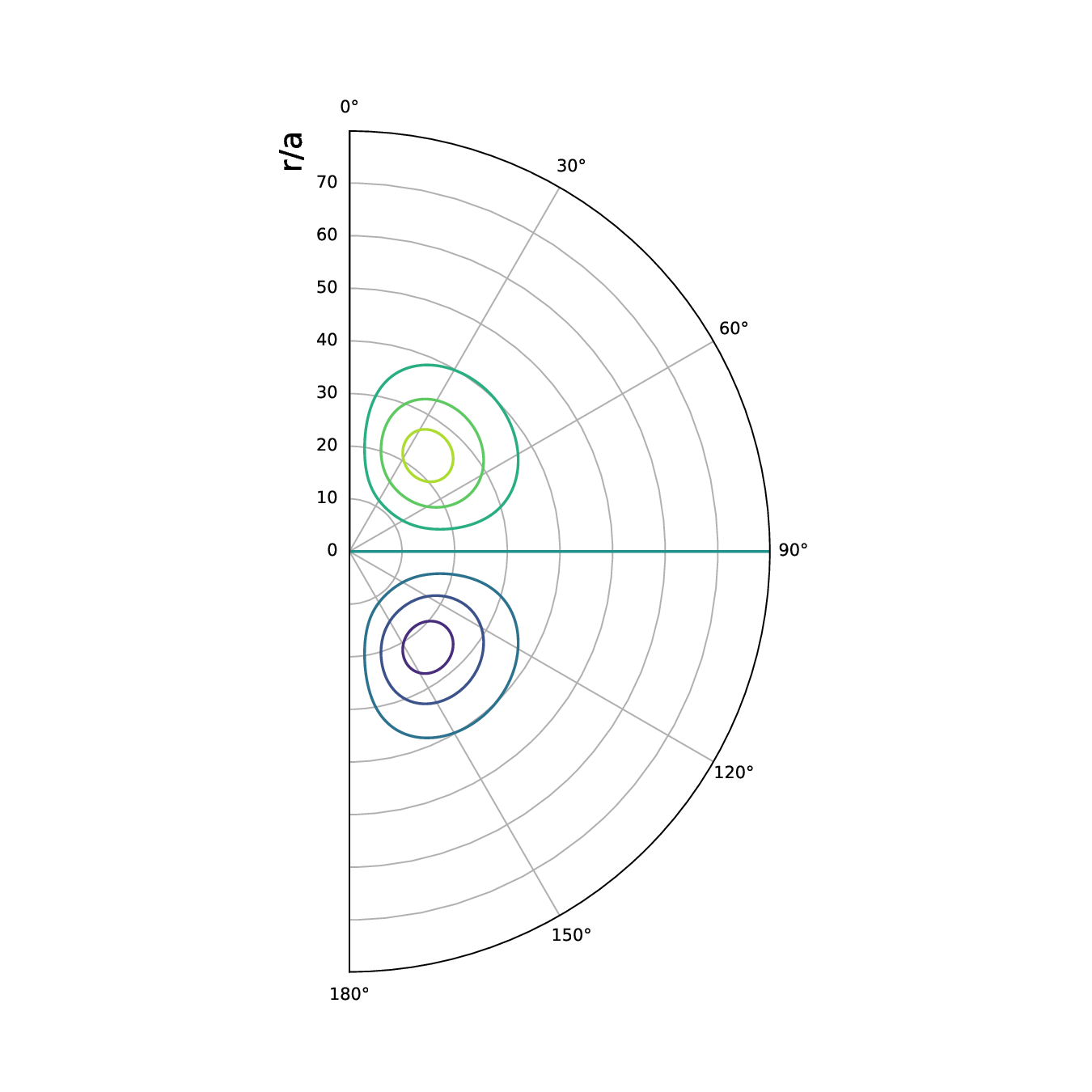} &
\includegraphics[width=8cm]{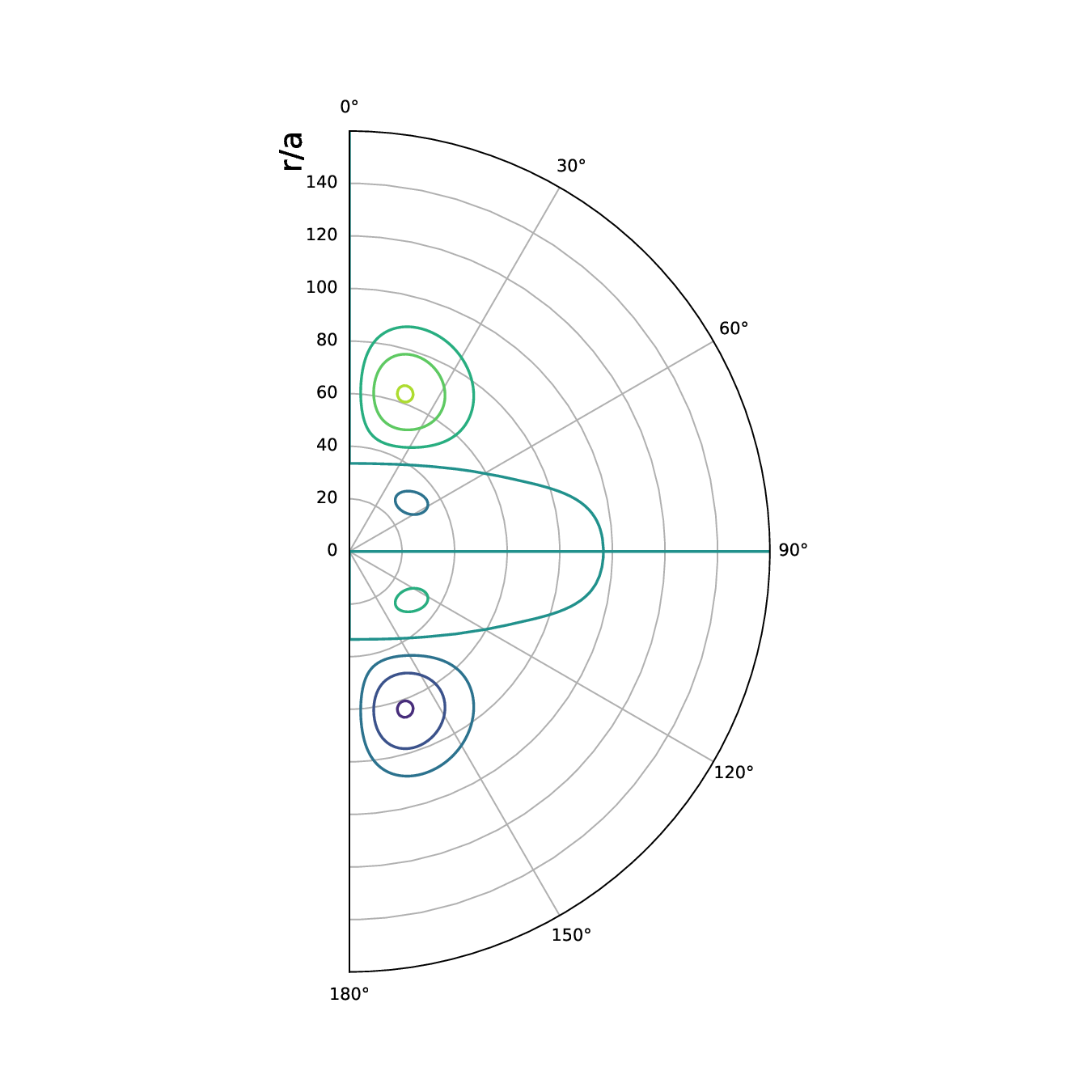} \\
(c) & (d)
\end{tabular}
\caption{\label{fig:ml1}
Probability densities for the four lowest-lying states
with $m_l=1$.  The associated $z$ parities and 
eigenenergies in units of $G^2m^5$ are 
(a) even, -0.05710;
(b) even, -0.01928;
(c) odd, -0.02900; and
(d) odd, -0.01402.
}
\end{figure}

\begin{figure}[ht]
\vspace{0.2in}
\begin{tabular}{cc}
\includegraphics[width=8cm]{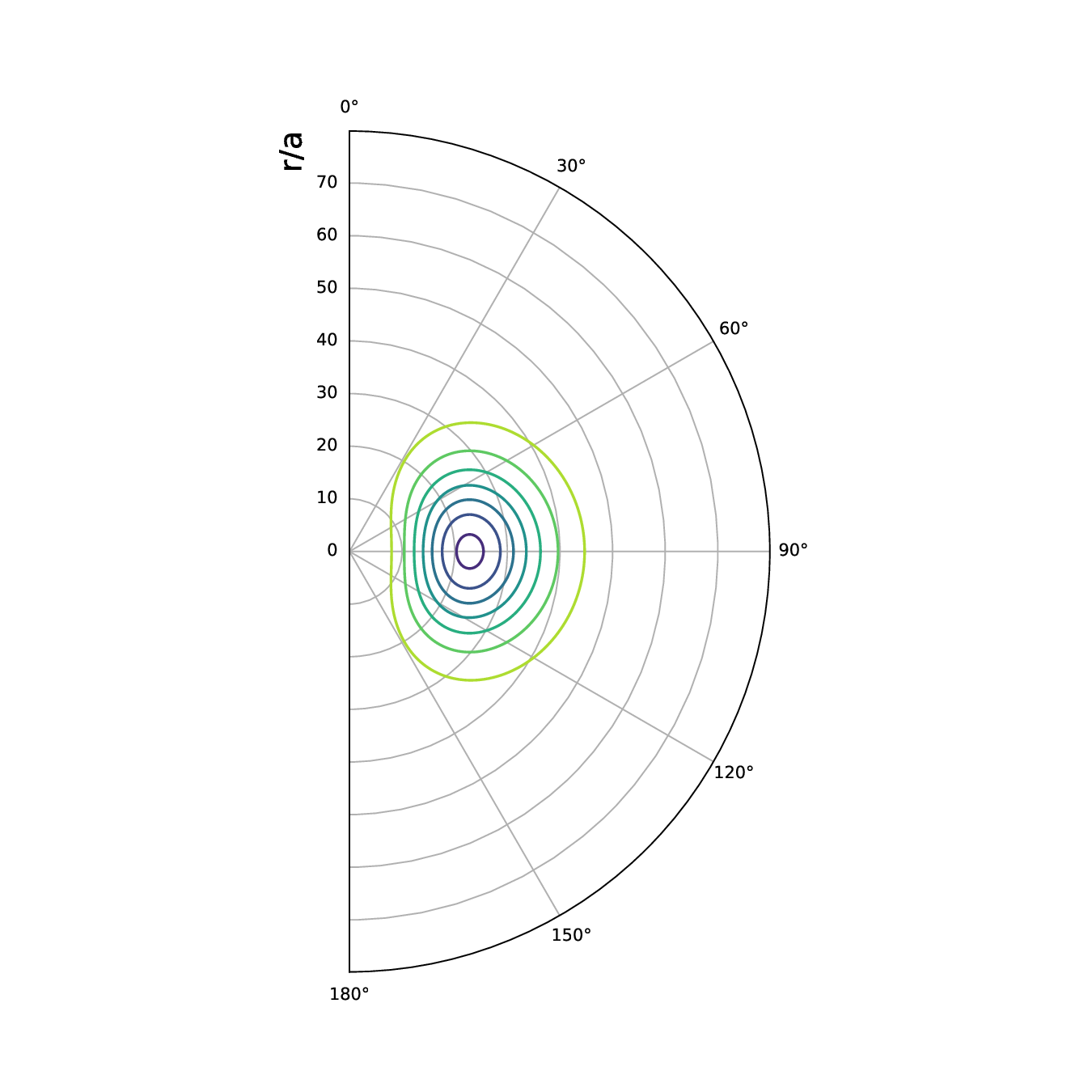} & 
\includegraphics[width=8cm]{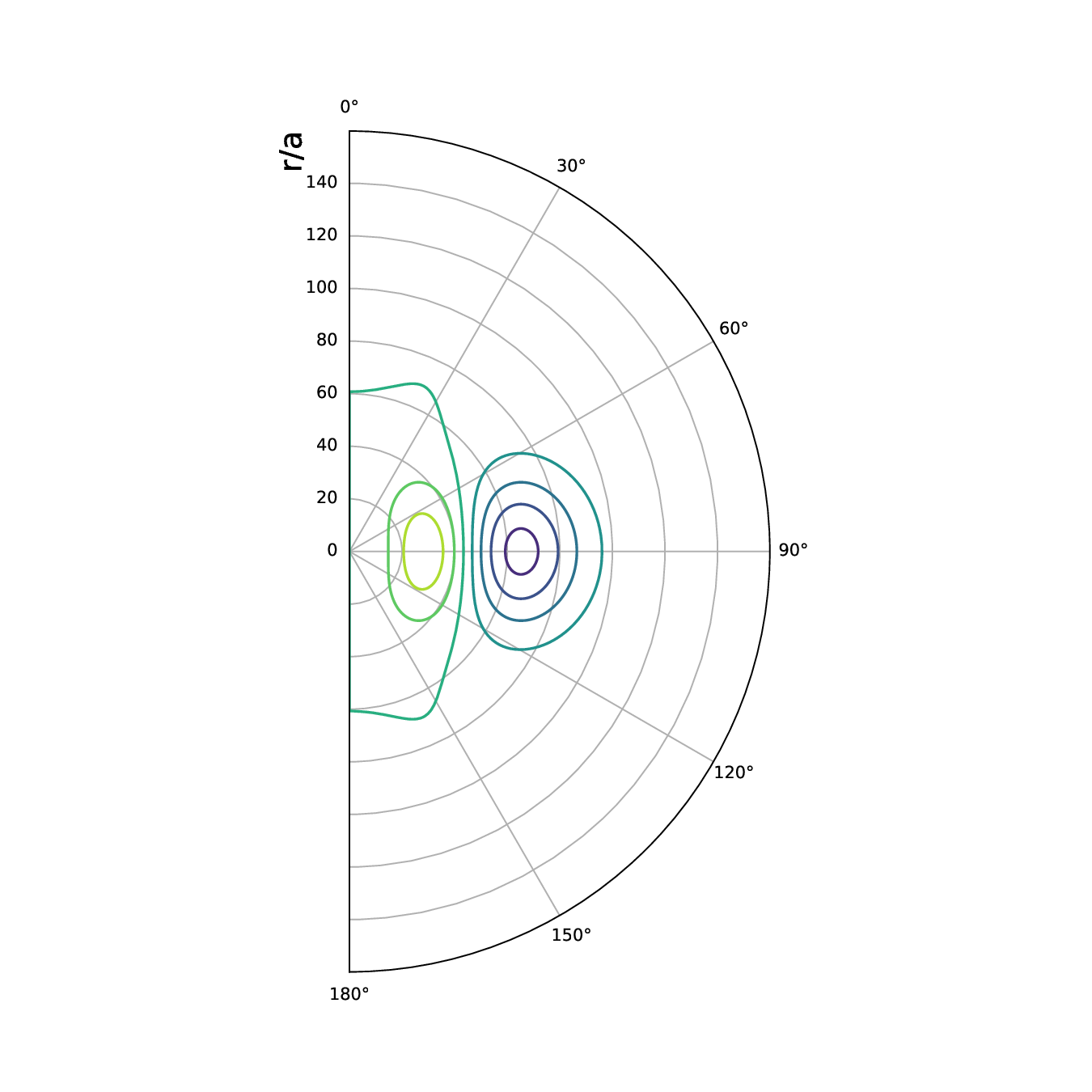} \\
(a) & (b) \\
\multicolumn{2}{c}{\includegraphics[width=8cm]{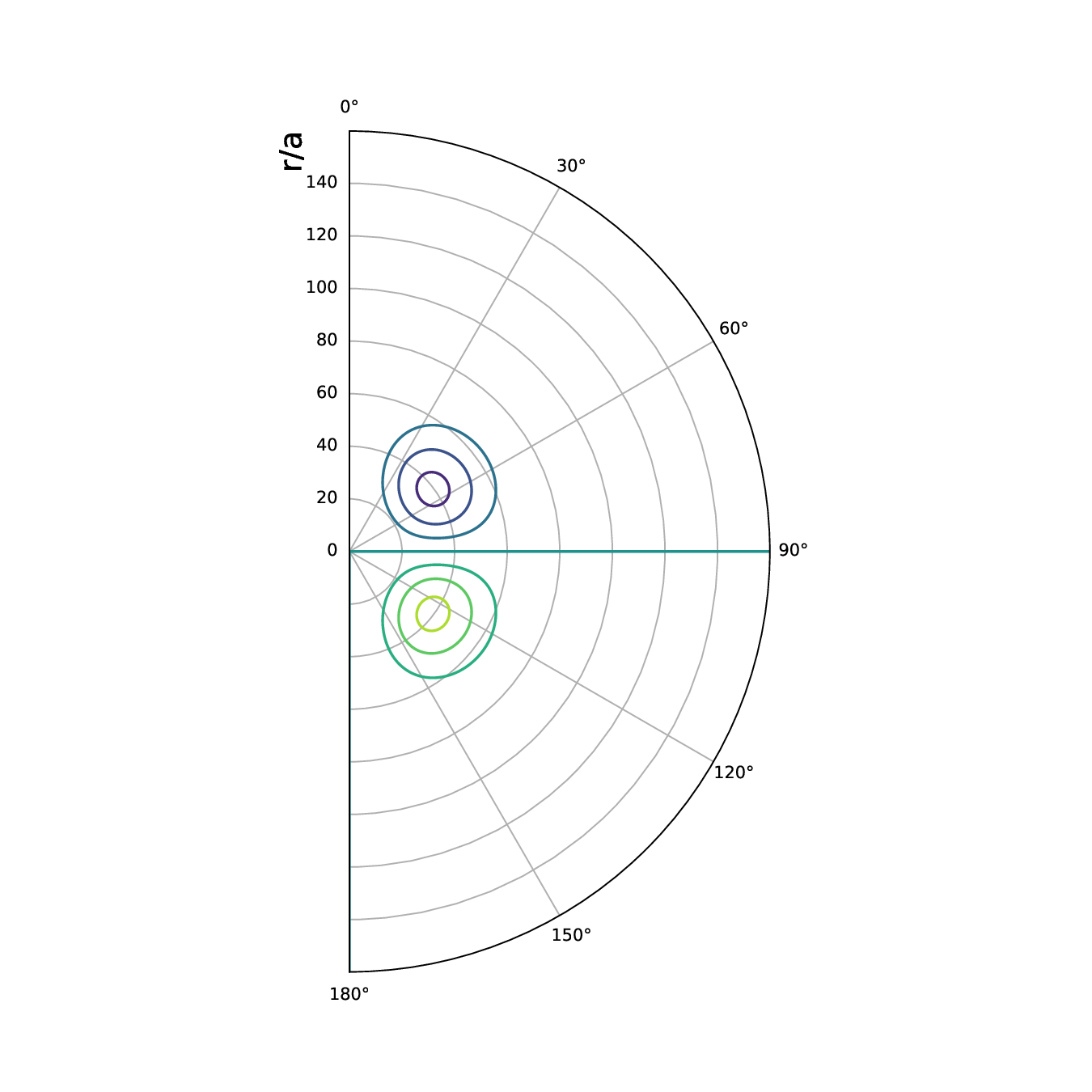}} \\
\multicolumn{2}{c}{(c)}.
\end{tabular}
\caption{\label{fig:ml2}
Probability densities for the three lowest-lying states
with $m_l=2$.  The associated $z$ parities and 
eigenenergies in units of $G^2m^5$ are 
(a) even, -0.03066;
(b) even, -0.01312; and
(c) odd, -0.01712.
}
\end{figure}

We have used these methods to compute some low-lying eigenstates
for $m_l$ values 0, 1, and 2 and for both $z$ parities.  The
energies are listed in Table~\ref{tab:eigenvalues}.  The relative
probabilities of the partial waves are presented in Tables~\ref{tab:evenell} and \ref{tab:oddell}.
States with energies outside the range of the energies in
the first column of Table~\ref{tab:eigenvalues} certainly exist but
were not considered.  The probability
densities obtained in each case are plotted in Figs.~\ref{fig:ml0even}-\ref{fig:ml2}.
As expected, an increase in the energy or the $m_l$ value brings
an increase in the complexity of the probability distributions.

\begin{table}[ht]
\caption{\label{tab:eigenvalues}
Eigenenergies in units of $G^2m^5$ for low-lying states
at fixed $m_l$ and $z$-parity even or odd, computed on
a grid from 0 to $r_{\rm max}$ with step size $h$ in a 
partial-wave expansion truncated at $l_{\rm max}$.
Typical values for the computational parameters are
$80a$ to $160a$ for $r_{\rm max}$ and $l_{\rm max}=10$,
with $h$ small enough to achieve four significant figures.
Only those states that fall within the energy range of 
the $m_l=0$, $z$-even subset are tabulated.
}
\begin{center}
\begin{tabular}{cc|cc|cc}
\hline \hline
\multicolumn{2}{c|}{$m_l=0$} & \multicolumn{2}{c|}{$m_l=1$} & \multicolumn{2}{c}{$m_l=2$} \\ 
even & odd & even & odd & even & odd \\
\hline
-0.1628  & -0.06894 & -0.05710 & -0.02900 & -0.03066 & -0.01712\\
-0.03082 & -0.0274  & -0.01928 & -0.01402 & -0.01312 & \\
-0.0252 & -0.0169  &          &          &          & \\
-0.01254 &          &          &          &          & \\
\hline \hline
\end{tabular}
\end{center}
\end{table}

\begin{table}[ht]
\caption{\label{tab:evenell}
Partial-wave content of low-lying states with contributions from even $l$ values, for various $m_l$
values and $z$ parities.  The eigenenergies $\epsilon_{m_l}$ are in units of $G^2m^5$.
The $m_l$ value and $z$ parity determine the contributing partial
waves as those with $l-m_l$ even (odd) for $z$ parity even (odd).
}
\begin{center}
\begin{tabular}{ccccccccc}
\hline \hline
      &            &                  & \multicolumn{5}{c}{$l$ for partial wave} \\
$m_l$ & $z$-parity & $\epsilon_{m_l}$ & 0  & 2 & 4  & 6  & 8 & 10\\
\hline
0 & even & -0.1628  & 1 &  &  &  & & \\
0 & even & -0.03082 & 1 &  &  &  & & \\
0 & even & -0.0252  & 0.7128 & 0.2576 & 0.02731 & 0.002217 & $6.13\times10^{-5}$ & $1.18\times10^{-6}$\\
0 & even & -0.01254 & 1 &  &  &  & & \\
1 & odd  & -0.02900 &  & 0.9837 & 0.01605 & 0.0002743 & $1.95\times10^{-5}$ & $4.0\times10^{-9}$ \\
1 & odd  & -0.01402 &  & 0.4505 & 0.4559 & 0.08598 & 0.007410 & 0.000240 \\
2 & even & -0.03066 &  & 0.9794 & 0.02021 & 0.0004160 & $7.82\times10^{-6}$ &  $1.37\times10^{-7}$ \\
2 & even & -0.01312 &  & 0.8752 & 0.1125 & 0.01128 & 0.0009063 & $6.51\times10^{-5}$ \\
\hline \hline
\end{tabular}
\end{center}
\end{table}

\begin{table}[ht]
\caption{\label{tab:oddell}
Same as for Table~\ref{tab:evenell} but for states with contributions from odd $l$ values.
}
\begin{center}
\begin{tabular}{cccccccc}
\hline \hline
      &            &                  & \multicolumn{5}{c}{$l$ for partial wave} \\
$m_l$ & $z$-parity & $\epsilon_{m_l}$ & 1 & 3 & 5 & 7 & 9\\
\hline
0 & odd  & -0.06894 & 0.9363 & 0.06193 & 0.001765 & $3.63\times10^{-5}$ & $6.34\times10^{-7}$ \\
0 & odd  & -0.0274  & 0.5940 & 0.3306 & 0.06479 & 0.009440 & 0.001250 \\
0 & odd  & -0.0169  & 0.6892 & 0.3000 & 0.01040 & $3.000\times10^{-4}$ & $1.23\times10^{-4}$ \\
1 & even & -0.05710 &  0.9936 & 0.006389 & $2.47\times10^{-5}$ & $7.46\times10^{-8}$ & $2.0\times10^{-10}$ \\
1 & even & -0.01928 &  0.8531 & 0.1376 & 0.008862 & 0.0004453 & $1.99\times10^{-5}$ \\
2 & odd  & -0.01712 &  & 0.9978 & 0.0007470 & 0.001440 & $8.4\times10^{-6}$ \\
\hline \hline
\end{tabular}
\end{center}
\end{table}

The eigenenergies of the spherically symmetric states agree 
with those calculated previously~\cite{Taylor}.  These are
the first, second and fourth energies listed in the first
column of Table~\ref{tab:eigenvalues}.  A state with only
axial symmetry has appeared in that column as the third 
entry; Table~\ref{tab:evenell} shows its partial-wave content,
with a 26\% probability for $l=2$.

One interesting aspect of the spectrum is that the first
excited state is not spherically symmetric.  It is instead
odd in $z$ with an energy of -0.06894~G$^2$m$^5$ and $m_l=0$.
Only odd-$l$ partial waves contribute, beginning with $l=1$,
and with a 6.2\% probability for $l=3$.  The probability
density is plotted in Fig.~\ref{fig:ml0odd}(a).  The energy of
this state agrees with the result
obtained by Schupp and van der Bij~\cite{Schupp}.\footnote{One
must take into account their different definition of a 
dimensionless energy $\hat{E}=-\epsilon/2$.}

The state with perhaps the most interesting structure
is the odd $m_l=1$ state depicted in Fig.~\ref{fig:ml1}(d)
with an energy of -0.01402~G$^2$m$^5$. The probability
density has four peaks and a partial-wave content of
45\% for $l=2$, 46\% for $l=4$, and 8.6\% for $l=6$.
It is the only state, of those computed, where the
second partial wave is as important as the first.  It
is also an example of why restriction to a single
partial wave, as done in \cite{Silveira}, will not
always provide a good approximation.

Any state with a partial-wave content for $l\geq2$
can transition to a lower state by emitting a
gravitational wave, which has helicity $\pm2$.  Thus many of the computed states
are actually unstable to radiation~\cite{Ferrell}.  Lifetimes for these states
could be computed, but that is beyond the scope of
the present work.

\section{Summary}  \label{sec:summary}

We have solved the Schr\"odinger-Newton problem for solitons
with axial symmetry using partial-wave expansions that show
the range of angular momentum content.  The energies obtained
are listed in Table~\ref{tab:eigenvalues} and include several
spherically symmetric states previously computed.  The
partial-wave content of each state is listed in Table~\ref{tab:evenell}
or \ref{tab:oddell}, depending on whether the content is even
or odd in $l$, respectively.  Plots of the mass distributions
are given in Figs.~\ref{fig:ml0even}-\ref{fig:ml2} and
show significant structure for those lacking spherical symmetry.
Some interesting aspects of the spectrum and states are
the presence of a nonspherical solution as the third 
state with azimuthal quantum number $m_l=0$, the first excited state being one
without spherical symmetry, and a state ($E=-0.01402$~G$^2$m$^5$)
with nearly equal contributions from $l=2$ and $l=4$ partial waves.

The methods employed provide a more complete picture of
these solitons than previously obtained, in that the 
sum over partial waves includes any with significant
contribution rather than approximating with a single
partial wave~\cite{Silveira}.  The calculations also
extend the study of axially symmetric solitons to cases
where the effective potential is not spherically 
symmetric, as is typically arranged for the study
of $l$-boson stars~\cite{Alcubierre,Jaramillo,Nambo}.

\acknowledgments
This work was supported in part by 
the Minnesota Supercomputing Institute 
and the Research Computing and Data Services
at the University of Idaho through
grants of computing time.

\appendix





\end{document}